\documentstyle[12pt]{article}

\parskip=\medskipamount                 
\def\7#1#2{\mathop{\null#2}\limits^{#1}}        
\def\greaterthansquiggle{\raise.3ex\hbox{$>$\kern-.75em\lower1ex\hbox{$\sim$}}}
\def\lessthansquiggle{\raise.3ex\hbox{$<$\kern-.75em\lower1ex\hbox{$\sim$}}}
\def\Nsquiggle{\raise.3ex\hbox{$N$\kern-.75em\lower1ex\hbox{$\,\sim$}}}

\begin{document}
\def\thefootnote{\fnsymbol{footnote}}
\begin{center}
{~~~}\\[.2in]
{\large\bf Threshold Effects on the
Mass Scale Predictions in SO(10) Models and Solar Neutrino
Puzzle}\\[.5in]
{\bf  R.N. Mohapatra and M.K.Parida\footnote{On leave from North-eastern
Hill University,Shillong,India}}\\[.2in]
{\it Department of Physics,University of Maryland,College Park,Md.20742}\\
{UMD-PP-92-179}\\
{March,1992}\\
\end{center}
\begin{center}
{\bf Abstract}\\
\end{center}

           We compute the threshold uncertainties due to
unknown masses of the Higgs bosons
on the predictions for the intermediate  and unification scales,
$M_I$  and  $M_U$  respectively
in SO(10) models.We focus on models with separate  breaking
scales for Parity and $SU(2)_R$ symmetries since they provide
a natural realization of the see-saw mechanism for neutrino
masses.
For the  two step symmetry breaking chains ,where
 left-right symmetric gauge groups
appear at the intermediate scale,
we find that parity invariance of the theory at the unification scale
drastically reduces the GUT threshold effects in some cases.
Including the effects of the intermediate scale
thresholds ,we compute the  uncertainty in the above mass scales
and study their implications for proton lifetime and neutrino masses.
An important outcome of our analysis is that if the currently
favored nonadiabatic MSW solution to the solar neutrino puzzle
is accepted , it will rule out the $SU(2)_LXSU(2)_RXU(1)_{B-L}X
SU(3)_c$ as an intermediate symmetry for SO(10) breaking
whereas the intermediate symmetry $SU(2)_LXSU(2)_RXSU(4)_c$,
is quite consistent with it.

\newpage
{\bf I.Introduction:}

The grandunified theories $^1$(GUTs)provide an elegant extension
of physics beyond the standard model.The  requirement that
the gauge couplings constants in these theories become
equal at the GUT scale ($M_U$) lends them a predictive power
which makes it possible to test them in experiments such as those
looking for the decay of the proton.The most predictive  such
theory is the minimal SU(5) model of Georgi and Glashow$^1$,where
the SU(5) symmetry breaks in one step to the standard model.The
only new mass scale in this model is the $M_U$ which can be
determined by the unification requirement using the low energy
values of any two gauge
couplings from the standard model.One then predicts not only
$M_U$, but also the remaining low energy gauge coupling constant
(say $sin^2\theta_W$).It is well known that for the minimal
SU(5) model,they lead to predictions for the proton life-time
as well as $sin^2\theta_W$ both of which are inconsistent
with experiments.

This however does not invalidate the idea of grandunification
and attention has rightly been focussed on SO(10) $^2$ GUT
models which can accomodate more than one new mass scale.
Supersymmetric SU(5) $^3$ models also belong to this class.
In this class of two mass scale thoeories,the values of low
energy gauge coupling constants can determine both the mass
scales again making these theories experimentally
testable.The determination
of the values of the new mass scales become more precise
as the low energy values of the gauge coupling constants
become better known.It is therefore not surprising that
the recent high precision measurement of $\alpha_{strong}$ and
$sin^2\theta_W$ at LEP$^4$  once again revived interest in
grandunified theories$^5$.

Supersymmetric SU(5) theories have
been studied with the goal of predicting the scale of
supersymmetry breaking$^5$.These models however donot have
any room for a nonzero neutrino mass nor natural generation
of adequate baryon asymmetry,whereas, the SO(10) model is
the minimal GUT scheme that provides a frame-work for a proper
understanding  both these problems.In this paper,we concentrate
on the SO(10) models with a two step breaking to the standard
model and study the threshold effects on the predictions
for the two new mass scales i.e. $M_U$ and $M_I$.In order to
appreciate the significance of our work, it is worth pointing
out that in SO(10) models the scale $M_U$ as usual is related
to proton decay whereas the intermediate scale is related
to neutrino masses if the intermediate symmetry is either
of the left-right symmetric $^6$groups $SU(2)_LXSU(2)_RXG_c$
where $G_c$ is $SU(4)_C$ or $SU(3)_cXU(1)_{B-L}$.If the neutrino
mass is determined independently (say, from the MSW solution to
the solar neutrino puzzle), then  the see-saw mechanism determines
the range of the required intermediate mass scale.The viability
of a given SO(10) model will then depend on both the  value
of $M_I$ obtained from renormalization group analysis as well
as the uncertainties in this value arising from threshold
corrections.

     Let us discuss the kind of SO(10) models we will study here.
As is wellknown,the SO(10) group contains the maximal subgroup
$SU(2)_LXSU(2)_RXSU(4)_c$xD, D being a
$Z_2$ symmetry which implements the parity
transformation(as well as particle-anti-particle transformation).
We will refer to this symmetry as D-parity. The
actual nature of SO(10) model depends on what symmetry
appears at the intermediate scales (i.e.
between the GUT scale and $M_W$).The most interesting SO(10) models
are the ones where the symmetry breaking to the standard model
occurs in two steps,with either of the left-right symmetric
groups $SU(2)_LXSU(2)_RXSU(4)_c$
or $SU(2)_LXSU(2)_RXU(1)_{B-L}XSU(3)_c$
(denoted henceforth by $G_{224} and G_{2213}$ respectively)
as the only intermediate symmetry.These are also the theories
for which definite predictions can be made.Our work will
focus on them.
Note the absence of D-parity at the intermediate scales .Use of Higgs
multiplets belonging to {45} and {210} representations to break SO(10) can
lead to such a scenerio as was pointed out in a series of papers
in 1984 by us in collaboration with D.Chang$^7$.Let us briefly
recapitulate some other motivation for considering such models.

One of the attractive features of the SO(10) models is that
they provide a natural understanding of the neutrino masses
via the see-saw mechanism$^8$.It has however been noted
that, the see-saw mass matrix does not follow naturally in
models where D-parity and $SU(2)_R$ breaking scales ($M_R$) are same.
On the other hand,if the D-parity breaking scale $M_P$ is such
that $M_P\gg M_R$ then the see-saw formula emerges naturally$^9$.
This is perhaps the most compelling motivation for requiring
the D-parity breaking scale to be significantly larger
than $M_R$.There
are however other motivations from cosmology.If $M_P$=$M_R$,
there arise domain walls bounded by strings at the epoch
when $SU(2)_RXD$ symmetry breaks down .They dominate the
mass density of the universe making it hard to understand
the successes of the big bang picture.Such problems do
not arise in SO(10) models with separate
D-parity breaking scenerios.Furthermore,  exact D-parity
leads to $n_B=n_{\bar{B}}$.In SO(10) models where the baryon
asymmetry of the universe arises from Higgs boson decays,
the ratio $n_B/n_\gamma$ receives an additional suppression
$\left(M_P/M_U\right)^2$ on top of its small value predicted
in generic GUT models.This mechanism would
prefer scenerios with D-parity breaking
scale to be above that of $SU(2)_R$  and at the GUT scale .
We are of course fully aware that,if baryon asymmetry
arises from the decay of heavy Majorana neutrinos,the above
constraint does not apply.

A complete two-loop analysis of the predictions for $sin^2\theta_W$
and proton life-time in this class of SO(10) models was carried out
in ref.10.Depending on the nature of the Higgs boson spectrum
used to implement the symmetry breaking and the nature of the
intermediate symmetry groups,the intermediate scales and the associated
physical implications were discussed. The mean values of
the mass scales will be taken from this paper.

A basic limitation of all grandunified theories is that all mass
scale predictions are subject to uncertainties arising from Higgs
boson thresholds$^{11}$.It has therefore been argued$^{12}$ that since
the Higgs bosons in question belong to large representations
in SO(10) theories, one might worry that the mass scale predictions
derived from two-loop calculations are completely unreliable.
In other words, even if $sin^2\theta_W$ and
$\alpha_{strong}$ are very precisely known,the unification scale
$M_U$ and the intermediate scale $M_I$ will have large uncertainties.
It was however subsequently pointed out that this need not
always be true;for instance, if an SO(10) model has an
intermediate symmetry group  $SU(2)_LXSU(2)_RXSU(4)_cXD$,
($G_{224D}$) the GUT threshold uncertainties in $sin^2\theta_W$
exactly cancel out$^{13}$.
This result of course holds only if the intermediate
symmetry is $G_{224D}$ and does not apply to
the more interesting models with separate D-parity breaking ;it also
does not say anything about the uncertainties due to intermediate
scale thresholds.It does however give rise to the hope that existence
of symmetries may reduce the net impact of threshold uncertainties.
In any case if the grandunified theories are to be useful,threshold
effects must be calculated.In this paper ,we begin this program
for the two SO(10) theories and hope to extend it to other models
later on.The main results of this paper have already been
reported earlier$^{14}$.

In ref.14 and in the present paper,we adopt the following approach.
Using the evolution equation for the coupling constants,
we express the $M_U$ and $M_I$ in terms of the known low energy
parameters $\alpha_{strong}$,$sin^2\theta_W$ and $\alpha_{em}$
and the threshold corrections due to Higgs bosons .Since the LEP
results have considerably reduced the experimental uncertainty
in $sin^2\theta_W$ as well as $\alpha_{s}$,the main uncertainty
comes from the arbitrariness associated with the Higgs boson
masses and the theoretical uncertainties in the scales
$M_U$ and $M_I$ can be computed.The final magnitude of the uncertainty depends
o
   n how far the
scalr masses are split from the symmetry breaking scale.Using the
standard model as a guide,we assume that constraints of one-loop
radiative corrections and unitarity bound on tree level amplitudes
would allow the scalar boson masses to be a factor of 10 on either
side of the symmetry breaking scale. We also present results for a
wider splitting of $(30)^{\pm1}$ for illustration even though we believe
this to be rather unlikely.

We consider the symmetry breaking chains:

(A)SO(10)$\rightarrow$$G_{224}$$\rightarrow$$G_{std}$;

 (B) SO(10)$\rightarrow$$G_{2213}$$\rightarrow$$G_{std}$ .

We compute the Higgs boson threshold effects on the uncertainty in
intermediate scale $(M_C/M^{0}_C)$ in case (A) to be
$10^{^{+2.7}_{-1.4}}$
and that in the value of grandunification scale i.e. in
$(M_U/M^{0}_U)$ to be $10^{^{+.8}_{-1.7}}$.This corresponds to a maximum
value for $\tau_p$ in case(a) to be $10^{40}$years for $\alpha_s$=.11
and $10^{38.4}$years for $\alpha_s$=.1.
For case (B),we find the uncertainty in $(M_R/M^{0}_R)$ to be
$10^{^{+.6}_{-.3}}$
and that in $(M_U/M^{0}_U)$ to be at most $10^{\pm.2}$.Note
that the threshold uncertainties are much less than the estimates
of ref.6.For $\alpha_s$=.11,we obtain an upper limit on $\tau_p$
in this case to be $5\times{10^{36}}$ years.

Using our results in combination with the see-saw formula
for neutrino masses,we find that the presently favored
nonadiabatic MSW solution to the solar neutrino puzzle
rules out the SO(10) model (B) which has
$SU(2)_LXSU(2)_RXU(1)_{B-L}XSU(3)_c$  as an
intermediate symmetry.In our opinion
this is an important result since this will be the second
GUT model  that is being definitively ruled out by experiment.
The symmetry breaking chain (model(A)) is however quite
consistent with data.
    This paper is organized as follows: in sec.2,we present a
derivation of the formulae for the threshold uncertainties
for the model (A);in sec.3,we derive the contribution of the
various Higgs multiplets to these uncertainties and give our
estimate of these effects.Sec.4,uses the results of sec.2 and 3
to derive the same results for the model(B);in sec.5, we derive
the implications of our results for the solar neutrino puzzle and
in sec.6,we discuss the implications of our results for
proton decay and the effect of adding extra Higgs multiplets
on our result.
In sec.7,
we close with some concluding remarks.

{\bf II.The Formula for the Threshold Corrections : model (A):}

Let us now proceed to derive the equations for the threshold
uncertainties.We will illustrate the technique for the model(A).
We start with the standard
renormalization group equations for the evolution of the
gauge coupling constants:

$$ \mu\left(d\alpha_i\over{d\mu}\right)={a_i\alpha_i^2\over{2\pi}}+
{1\over{8{\pi^2}}}\Sigma_j{b_{ij}
\alpha_i^2\alpha_j}   \;    \eqno (1)$$

In eq.(1),$\alpha_i={g_i^2\over4\pi}$ and the one loop coefficient
$a_i=-{11N\over3}+{4\over3}n_g +{T(s)\over3}$,where $n_g$ is the number
of fermion generations and T(s) is the contribution of the Higgs bosons.
The $b_{ij}$ are the two loop coefficients,which are not needed here.
At each symmetry breaking threshold,we use the following maching
conditions$^{14}$:
(we assume that the group $G_I$ breaks to the group $G_i$ at the
scale $M_I$)

$$ {1\over{\alpha_i(M_I)}}={1\over{\alpha_I}}-{\lambda^I_i\over{12\pi}}
                       \;              \eqno(2)   $$

In eq.(2), $\lambda^I_i=Tr(\theta_i^{V})^2+Tr(\theta_i^H)^2ln(M_H/M_I)$;
$\theta^V_i$ are the generators of the lower symmetry $G_i$ for the
representations in which the heavy Gauge bosons appear;$\theta_i^H$
is the same for the superheavy Higgs bosons.

Let us now apply the formulae in eq.(1) and (2) to the SO(10) models
described earlier as (A) .Note that since the D-parity has been broken
in both cases at the GUT scale,the Higgs multiplets needed to implement
the symmetry breaking in model (A) are in {210},{126} and{10} dimensional
representations of the SO(10) group.We denote by $M_U$,$M_C$ and $M_Z$ the
three symmetry breaking scales  and they arise from the vev's
of the above three Higgs multiplets respectively.
The D-parity breaking manifests itself in the mass of the submultiplet
of the {126} representation $\Delta_L$ (transforming as (3,1,10) under
$G_{224}$) being different from the submultiplet $\Delta_R$ transforming
as $(1,3,\bar{10})$ under the same group.We postpone any further
discussion of the Higgs bosons to the next section.Let us now derive
the formula for the threshold corrections.

Using eq(1) and (2) and the standard Georgi-Quinn-Weinberg type
analysis,we find the following expressions for $sin^2\theta_W$
and $\alpha_{s}$:

$$16\pi\left(\alpha_{s}^{-1}-{3\over8}\alpha_{em}^{-1}\right)=
A_cln\left(M_c\over{M_Z}\right)+A_Uln\left(M_U\over{M_Z}\right)
+F_c+F_{\lambda}+{\Gamma_{s}} \;\eqno(3)$$
where

$A_c=(8a_3-3a_{2L}-5a_Y-6a_4^{\prime}+3a_{2R}^{\prime}+
3a_{2L}^{\prime})$;

$A_U=(6a_4^{\prime}-3a_{2R}^{\prime}-3a_{2L}^{\prime})$;

$F_c=3ln\beta(a_{2L}^{\prime}-a_{2L}^{''}+a_{2R}^{'}-a_{2R}^{''}
-2a_{4c}^{'}+a_{4c}^{''})$;

$F_{\lambda}=-(4/3)(\lambda_{3c}^C-{3\over8}\lambda_{2L}^C-{5\over8}
\lambda_Y^C+{3\over4}\lambda_{4c}^U-{3\over8}\lambda_{2L}^U-
{3\over8}\lambda_{2R}^U)$.

$${{16\pi}\over{\alpha_{em}}}\left(sin^2\theta_W-{3\over8}\right)=
B_cln{M_c\over{M_Z}}+B_Uln{M_U\over{M_Z}}+G_c+G_{\lambda}
+\Gamma_{\theta} \;\eqno(4)$$

where

$B_c=(5(a_{2L}-a_Y)-(5a_{2L}^{'}-2a_{4c}^{'}-3a_{2R}^{'}))$;

$B_U=(5a_{2L}^{'}-2a_{4c}^{'}-3a_{2R}^{'})$;

$G_c=ln\beta(5a_{2L}^{''}-3a_{2R}^{''}-2a_{4c}^{''}-5a_{2L}^{'}
+3a_{2R}^{'}+2a_{4c}^{'})$;

$G_{\lambda}={5\over6}(\lambda_Y^C-\lambda_{2L}^C+{3\over5}\lambda
_{2R}^U+{2\over5}\lambda_{4c}^U-\lambda_{2L}^U)$

In the above expressions, $a_i$,$a_i'$ and $a_i''$ denote
the the evolution coefficient
for the gauge couplings between $M_W$ to$M_C$,$M_C$ to $M_P$ and $M_P$
to $M_U$ respectively.$M_P$ is the scale of D-parity breaking.
The $\Gamma$'s denote the two loop contributions, which do not
contribute to threshold uncertainties to the leading order
and will therefore be omitted henceforth.
The values of the $a_i$'s for model (A) are:
$a^{''}_{2L}=a^{''}_{2R}=a^{'}_{2R}=11/3,a^{'}_{2l}=-3,
a^{'}_{4c}=-23/3,a^{''}_{4c}=-14/3,a_{Y}=41/10,a_{3c}=-7;$

They will be used in the numerical estimates of the various
effects.
Using eq(3) and (4), we can express the mass scales $M_C$
and $M_U$ in terms of the low energy parameters and
the threshold contributions.The uncertainties in the
low energy parameters are experimental and  can be
estimated to be small as we show.The threshold contributions
buried in the $\lambda$'s intrduce the theoretical
uncertainty having to do with the fact that the heavy
Higgs masses are unknown.

Let us first address the uncertainties due to the experimental
errors in $sin^2\theta_W$ and $\alpha_{s}$ which we take $^{16}$
as:
$$sin^2\theta_W=.2334\pm.0008; \alpha_{s}=0.115\pm.007\;\eqno(5)$$

Denoting by $C_0={16\pi\over\alpha_{em}}({\alpha_{em}\over{\alpha_{s}}}-
{3\over8})$ and $C_1={16\pi\over\alpha_{em}}(sin^2\theta_W-{3\over8})$,
we get

$$(\Delta ln{M_C\over{M_Z}})_{expt}=
\Delta({{C_0B_U-C_1A_U}\over{A_cB_U-A_UB_c}})\;\eqno(6)$$

$$(\Delta{ln{M_U\over{M_Z}}})_{expt}=\Delta({{C_0 B_c-C_1 A_c}\over
{B_U A_c-A_U B_c}})\;\eqno(7)$$

Using eq.(3) and(4), and the values of various $a_i$'s
given above to evaluate the A's and B's ,
we can now estimate the uncertainty in the
quantities $M_C\over{M_C^0}$ and $M_U\over{M_U^0}$ where the
quantities with subscript denotes the values corresponding to the
mean values of $sin^2\theta_W$ and $\alpha_{s}$:

$${M_C\over{M_C^0}}=10^{\pm.025}; {M_U\over{M_U^0}}=10^{\pm.22}\;\eqno(8)$$

 We will find that these uncertainties are small compared to
those arising from unknown masses of heavy higgs bosons,thanks
to precision experiments from LEP $e^{+}e^{-}$ collider.

Again using eq.(3) and (4) and the expressions for A,B,F and G,
we can derive the following expressions for the uncertainties in
$M_U$ and $M_C$ arising from threshold effects only:

$$ \Delta{ln}\left({M_U}\over{M_Z}\right)= {f^P}_{M}+{f^U}_{M}+{f^C}_M
-({A_M/A_\theta})\left({f^P}_\theta+{f^U}_\theta+{f^C}_\theta\right)
+(\Delta{ln M_C})_{expt}(\alpha_{s},sin^2\theta_W)\; \eqno(8)$$

$$ \Delta{ln}\left({M_C}\over{M_Z}\right)=-{1\over{A_\theta}}({f_\theta}^P
+{f_\theta}^U+{f_\theta}^C) +(\Delta{ln M_U})_{expt}(\alpha_{s},sin^2\theta_W)
 \; \eqno(9)$$

where $$ {f^P}_{M}=\left(1-{{a_{2L}''+a_{2R}''-2a_{4c}''}
\over{a_{2L}'+a_{2R}'
-2a_{4c}'}}\right)ln\beta  \;\eqno(10a)$$

$${f^U}_{M}=\left({\lambda^U_{2L}-\lambda^U_{4c}}\over3( a_{2L}'
+a_{2R}'-2a_{4c}')\right)  \; \eqno(10b)$$

and $${f^C}_{M}=\left({5\lambda^C_Y+3\lambda^C_{2L}-8\lambda^C_{3c}}\over
18(a_{2L}'+a_{2R}'-2a_{4c}')\right)  \;\eqno(10c)$$

$${f^P}_{\theta}=-\left({(a_{4c}''-a_{2L}'')
(a_{2R}'-a_{2L}')ln\beta}
\over{(a_{2L}'+a_{2R}'-2a_{4c}')}\right) \;\eqno(10d)$$

$${f^U}_{\theta}=\left({(\lambda^U_{4C}-\lambda^U_{2L})(a_{2R}'
-a_{2L}')}\over{6(a_{2L}'+a_{2R}'-2a_{4c}')}\right)    \;\eqno(10e)$$

$${f^C}_{\theta}={1\over{6}}\left({\lambda^C_{2L}(a_{4c}'-a_{2R}')
+(5/3)\lambda^C_Y(a_{2L}'-a_{4c}')+\lambda^C_{3c}(a_{2R}'+(2/3)a_{4c}-(5/3)a_{2L
   }')}\over{a_{2L}'+a_{2R}'-2a_{4c}'}\right)..   \;\eqno(10f)$$

$${A_M}=1-\left({{5/3}a_Y+a_{2L}-{8/3}a_{3c}}\over{a^{'}_{2L}+a^{'}_{2R}
-2a^{'}_{4C}}\right)  \;  \eqno(10g)$$

$$A_\theta={5/8}\left(1\over{a^{'}_{2L}+a^{'}_{2R}-2a^{'}_{4C}}\right)B
\;\eqno(10h)$$

where $$ B=\left(({3/5}a^{'}_{2R}+{2/5}a^{'}_{4C}-a^{'}_{2L})({5/3}a_Y+a_{2L}
-{8/3}a_{3c})-(a^{'}_{2L}+a^{'}_{2R}-2a^{'}_{4C})(a_Y-a_{2L})\right)$$

In eq(10) ,$\beta=M_U/M_C$.
if the intermediate symmetry is $G_{224D}$,one can still use the above
general expressions,after dropping the $f^P_\theta$ and $f^P_M$ terms
in eq(8) and(9).(In this case of course,all $a^{'}=a^{''}$.) Note
that,in this case $f^U_\theta$ and $f^P_\theta$ vanish. To see
that this is what one expects from the results
of Parida and Patra(ref.13),
we note that in their work (ref.13),the uncertainty in $M_C$ was
assumed to be zero. Using this and bringing back the $sin^2\theta_W$
and $\alpha_{s}$ terms to the equation (9), $f^U_\theta$ and
$f^P_\theta$ terms  can be identified as the GUT threshold
 uncertainty in $sin^2\theta_W$ and therefore their vanishing
in the $G_{224D}$ limit was what was established in ref.12.

{\bf III. Survival Hypothesis and Estmation of the Threshold Uncertainties:}

In order to give a numerical estimate of these uncertainties,we need
to know the masses of the physical Higgs bosons;more specifically,
what submultiplets  are at what mass scale.This can be done using
the survival hypothesis for the Higgs bosons$^{17}$.The basic
assumption of the survival hypothesis is that only a minimal
number of fine tunings of the parameters in the Higgs potential
are done as required to ensure the hierarchy of the various
gauge boson masses.In the case at hand we need to fine tune only
two parameters since we have only a two step breaking.The survival
hypothesis then says $^{17}$that a submultiplet of the Higgs multiplet
of the GUT group,that acquires a vev to break a given subgroup
$G_i$ is stuck at the symmetry breaking scale.The other submultiplets
which transform as complete irreducible representations under $G_i$
get pushed to the next higher scale.Using this,we find the scales
for the Higgs boson masses .

For the case of model(A),the Higgs multiplets needed are
{210},{126} and two {10}-dimensional multiplets.They are
responsible for the three symmetry breaking scales $M_U$,
$M_I=M_C$ and $M_W$ respectively.As is well-known,the {210}
multiplet also breaks the D-parity symmetry.Using the survival
hypothesis,the scales of the different Higgs multiplets can
be obtained and they are listed in table I and II. In table I,
we list the Higgs bosons with masses around $M_U$;in Table II,
the Higgs bosons with masses near $M_I$ are listed.

 The U-submultiplet is the Goldstone mode corresponding
to the superheavy Gauge bosons and are omitted
 in computing the threshold uncertainties.

Let us now give their contributions to the various
$\lambda^U$'s and to final uncertainties.Defining
$\eta_i=(ln{{M_i}\over{M_U}})$, we find:

$$\lambda^U_{2L}=\lambda^U_{2R}=6+30\eta_{\zeta_0}+30\eta_{\Sigma}+
20\eta_{\zeta}   \;   \eqno(11)$$

Here,we have assumed that $M_{\Sigma_L}=M_{\Sigma_R}$ by left-right
symmetry.Similarly,$\zeta_1$ and $\zeta_2$ effects are combined and
denoted by $\zeta$.The coefficients in front of the different $\eta$'s
are simply the Dynkin indices of the different multiplets under the
different gauge groups.For instance,for $\eta_{2L}$,the Dynkin index
is that of the gauge group $SU(2)$ etc.

$$\lambda^U_{4C}=4+2\eta_H+2\eta_S+32\eta_{\zeta_0}+24\eta_\Sigma+
24\eta_{\zeta}+4\eta_{\zeta_3}  \;   \eqno(12)$$

Repeating the similar procedure for the intermediate scale,we find
the contribution to $\lambda^C$'s to be as follows:

$$\lambda^C_Y={1/5}\left(3\eta_{\phi}+2\eta_{R_1}+4\eta_{R_2}+16\eta_{R_3}
+32\eta_{R_4}+64\eta_{R_5}+24\eta_{R_6}+14\right)   \; \eqno(13)$$

$$\lambda^C_{2L}=\eta_\phi  \; \eqno(14)$$

$$\lambda^C_{3c}=1+\eta_{R_1}+5\eta_{R_2}+5\eta_{R_3}+\eta_{R_4}+5\eta_{R_5}
   \;\eqno(15)$$

The threshold contributions to $\Delta ln({M_C\over{M_Z}})$ and
$\Delta ln({M_U\over{M_Z}})$ can now be written down :

$$\Delta ln({M_C\over{M_Z}})=\left(1.29 ln\beta + .0259\left(-2+2\eta_{10}
+ 4\eta_{126} +2\eta_{210}\right)+.0017\left(65-15\eta_\phi +769\eta_R
\right)\right) \;\eqno(16)$$

$$\Delta ln({M_U\over{M_Z}})=-.92ln\beta-.0621\eta_{10}-.124\eta_{126}
-.062\eta_{210}+.031\eta_{\phi}-.496\eta_{R}\;\eqno(17)$$

In deriving eq(16) and(17),we have assumed that
the Higgs multiplets belonging to a single SO(10) supermultiplet have
the same mass at a given scale$^{18}$. This implies that $\eta_{\zeta_{0}}
=\eta_{S}=\eta_{126}$ , $\eta_{\Sigma}=\eta_{\zeta}$ and all $\eta_{R_{i}}$
are equal.
While there can be deviations
from this degeneracy assumption,their contributions will not change
our results noticeably.Secondly,we allow the $e^{\eta}$'s
to be between (1/10)and (10) as well as 1/30 and 30.  The first case
corresponds
to allowing the Higgs self-scalar couplings to range from $10^{-2}$
to $10^{2}$ times the gauge coupling and is better motivated by the
analogy of standard model than the second choice, although,we
present the results for both cases. As mentioned above the value of
$M_{\Delta_L}$is always kept lower than $M_U$,which corresponds to
taking $\beta$ less than one.Using eq(16) and(17),
 we compute the  uncertainties in the intermediate
mass scales as well as the $M_U$.We present these results in table III.
There are two possibilities;one when the uncertaintyin $M_C$ is
maximized and another when the uncertainty in $M_U$ is maximized.
We find the maximal uncertainty in $M_U\over{M_U^0}$ to be
$10^{^{+.8}_{-1.7}}$ whereas that in $M_C\over{M_C^0}$ to be
$10^{^{+2.8}_{-1.5}}$ from Higgs boson threshold effects alone.
We will study the impact of our results on the predictions of the
SO(10) model in sec.5.

{\bf IV.Threshold Corrections for Model (B):}

Let us now turn our attention to discussing model(B),where the SO(10)
symmetry first breaks down at the scale $M_U$ to $SU(2)_LXSU(2)_RX
U(1)_{B-L}XSU(3)_c$ which subsequently breaks down at scale $M_R$
to the standard model.Again as before the D-parity symmetry is
broken at the GUT scale.The Higgs multiplets necessary to implement
this chain are {45},{54},{126} and {10} dimensional ones.The D-
Bodd component of the {45}-dim. Higgs breaks the GUT symmetry.It has
been pointed out that $^{19}$, without the presence of {54},{45}
will break SO(10) down to SU(5)XU(1) rather than $G_{2231}$.However
as far as the threshold corrections are concerned,in the limit
of exact degeneracy the contribution of the submultiplets of
{54} exactly cancel.
Let us now present the equations for the
threshold corrections to $M_U$ and $M_R$ in this case.
Using the same notation as in eq.(8) and (9),(except
that we replace $M_C$ by $M_R$ in eq(9)),we give the
expressions for the various f's and A's below.
Defining $A^{'}_U=8a^{'}_{3c}-3a_{2L}^{'}-3a_{2R}^{'}-2a_{BL}^{'}$
and $B^{'}_U=5a_{2L}^{'}-3a_{2R}^{'}-2a_{BL}^{'}$,we get

$$A_{\theta}={1\over A_U^{'}}(8a_{3c}-3a_{2L}-5a_{Y}-{5\over B_U^{'}}
(a_{2L}-a_Y)) \;\eqno(18a)$$

$$A_M=\left(1-{1\over A_U^{'}}(8a_{3c}-3a_{2L}-5a_Y)\right)\;
\eqno(18b)$$

$$f_\theta^P=ln\beta\left({1\over{A_U^{'}}}(8a_{3c}^{''}-6a_{2L}^{''}
-2a_{BL}^{''})-{2\over{B_U^{'}}}(a_{2L}^{''}-a_{BL}^{''})\right)\;
\eqno(18c)$$

$$f_\theta^U={-1\over{6 A_U^{'}}}(8\lambda_{3c}^U-6\lambda_{2L}^U
-2\lambda_{BL}^U)+{2\over{6 B_U^{'}}}(\lambda_{2L}^U-\lambda_{BL}^U)\;
\eqno(18d)$$

$$f_\theta^C={-1\over{6 A_U^{'}}}(8\lambda_{3c}^R-3\lambda_{2l}^R
-5\lambda_Y^R)+{5\over{6 B_U^{'}}}(\lambda_{2L}^R-\lambda_Y^R)\
;\eqno(18e)$$
$$f_M^P=ln\beta\left(1-{1\over{A_U^{'}}}(8a_{3c}^{''}-6a_{2L}^{''}
-2a_{BL}^{''}\right) \;\eqno(18f)$$

$$f_M^U={1\over{6A_U^{'}}}(8\lambda_{3c}^U-6\lambda_{2L}^U-2\lambda_{BL}^U)
\;\eqno(18g)$$

$$f_M^C={1\over{6A_U^{'}}}(8\lambda_{3c}^R-3\lambda_{2L}^R-
5\lambda_Y^R)\;\eqno(18h)$$

In order to evaluate the threshold corrections,we need the
values of gauge coupling evolution coefficients $a_i$'s as
well the mass scales of the physical Higgs bosons.
We have $a_{2L}^{''}=a_{2R}^{''}=a_{2R}^{'}=-7/3$,$A_{BL}^{''}=7$
$a_{3c}^{'}=a_{3c}^{''}=-7$,$a_{2L}^{'}=-3$,$ a_{BL}^{'}=11/2$
$a_{3c}=-7;a_{2L}=-19/6,a_Y=41/10$.

Next we will use the survival hypothesis to determine the mass
scales of the various Higgs bosons.In table IV,we give the
Higgs bosons with masses of order of$M_U$.We do not include
the components of {54}-dim multiplet since in the degenerate
multiplet approximation ,their effects cancel out exactly.
The Higgs fields
with masses of order $M_R$ are only two in number and are therefore
listed in the text.

There are only two Higge multiplets at scale $M_R$; they are
$\Delta_R^{++}(1,2\sqrt{3/5},1)$ and $\phi(2,-{1\over{2}}\sqrt{3/5},1)$
where the numbers within the bracket refer to their transformation
property under the standard model.
 Their contribution to the various $\lambda$'s are given below:

$$\lambda_{2L}^U(126)=24\eta_{H_{1L}}+12\eta_{H_{2L}}+6\eta_{H_3}
+6\eta_{H_4}+16\eta_{H_5}+2\eta_{H_6}\;\eqno(19a)$$

$$\lambda_{2R}^U(126)=\lambda_{2L}^U(126)$$

$$\lambda_{3c}^U(126)=30\eta_{H_1}+6\eta_{H_2}+4\eta_{H_3}+4\eta_{H_4}
+24\eta_{H_5}+\eta_{H_7}+\eta_{H_8}  \;\eqno(19b)$$

$$\lambda_{BL}^U(126)=12\eta_{H_1}+6\eta_{H_2}+16\eta_{H_3}+16\eta_{H_4}
+\eta_{H_7}+\eta_{H_8}  \;\eqno(19c)$$

$$\lambda_{2L}^U(10)=\lambda_{2R}^U(10)=0  \;\eqno(19d)$$

$$\lambda_{BL}^U(10)=\lambda_{3c}^U(10)={1\over2}(\eta_{T_1}
+\eta_{T_2}) \;\eqno(19e)$$

$$\lambda_{2L}^U(45)=2\eta_{S_2} \;\eqno(19f)$$
$$\lambda_{2R}^U(45)=2\eta_{S_3} \;\eqno(19g)$$

$$\lambda_{3c}^U(45)=3\eta_{S_1}  \; \eqno(19h)$$

$$\lambda_{BL}^U(45)=0  \;\eqno(19j)$$

$$\lambda_Y^R=6/5+24/5\eta_{R_1}+{3\over5}\eta_{\phi}  \;\eqno(19k)$$

$$\lambda_{2L}^R=\eta_{\phi}   \;\eqno(19l)$$

$$\lambda_{3c}^R=o  \;\eqno(19m)$$

Using eq(18) and (19) and the values of $a_i$'s,we get for the
threshold contribution to the uncertainties in $M_U$ and $M_R$
the following expressions:

$$\Delta{ln({M_U\over{M_Z}})}=.0685ln\beta-.171\eta_{126}-.049\eta_{45}
-.039\eta_{10}-.177\eta_{\phi}-.146\eta_{R_1}  \;\eqno(20)$$

$$\Delta{ln({M_R\over{M_Z}})}=.095ln\beta-.083\eta_{126}+.033\eta_{45}
+.062\eta_{10}-.06\eta_{\phi}+.22\eta_{R_1}  \;\eqno(21)$$

In order to evaluate the possible uncertainties,as before we keep
$\beta$ less than one and allow $e^{\eta}$ to vary between .1 to
10 in one case and 1/30 to 30 in the second case.The results are
given in table III.WE see that in this case the threshold
uncertainties are much less than in model (A). The maximal
uncertainty in $M_R\over{M_R^0}$ is $10^{^{+.6}_{-.3}}$
whereas that in $M_U\over{M_U^0}$ is $10^{\pm.2}$.This has

We also wish to note at this point the uncertainties in $M_U$
and $M_R$ arising from the errors in $\alpha_{s}$ and $sin^2\theta_W$:
Using the same formula as in eq (6) and (7), we find for model (B):
$${{M_U\over{M_U^0}}=10^{\pm.25}};{{M_R\over{M_R^0}}=10^{\pm.18}}\;
\eqno(22)$$

{\bf V.Solar Neutrino Puzzle and SO(10):}

In this section,we study the implications of the results derived in
this paper for solar neutrino puzzle.As is well known,one of the
most interesting resolutions of the solar neutrino deficit is
the so called MSW matter oscillation mechanism$^{20}$.In this
mechanism,resonant enhancement of the oscillation of $\nu_e$
to either $\nu_{\mu}$ or $\nu_{\tau}$ takes place in the
solar core for a range of values of the ${\Delta{m}}^2$ and
the mixing angle $\theta$.In the so-called high mass (adiabatic)
solution, the value of $\Delta{m}^2$ is of order $10^{-4} {ev^2}$
 with $sin^2\theta\simeq{0.02 -.6}$
whereas in the nonadiabatic solution,we instead have
$\Delta{m}^2 sin^2{2\theta}\simeq {4X10^{-8}}{eV^2}$ with
${\Delta{m}}^2\simeq{10^{-6}{eV^2} - 8\times10^{-8}{eV^2}}$.
The combination of Chlorine$^{21}$ and Kamiokande $^{22}$
and initial Gallium data $^{23}$
 seems to point towards the nonadiabatic solution $^{20}$.
Either of the cases seem to fit quite well with the see-saw
picture for the neutrino masses in the SO(10) model $^{24}$
 provided
one assumes D-parity breaking$^9$.In the presence of D-parity
breaking,the hierarchical quadratic mass formula for neutrino
masses follows naturally ( a fact,which appears not to have
been well appreciated by many theorists).The point,briefly
is that,due to the presence of couplings of Higgs bosons
of type 126.126.10.10,a vev of $\Delta_R$ induces a vev
of the {126} submultiplet $\Delta_L$ of order $m_W^2\over {v_R}$
.This leads to a direct mass for all left-handed neutrinos
of the same order invalidating the conventional
 see-saw mechanism formulae.If however,the D-parity is broken
at the GUT scale,the $\Delta_L$ vev becomes only of order$^9$
$m_W^2{v_R}\over{M_U^2}$ which is smaller than the see-saw
contribution to the neutrino masses.

After the radiative
corrections are taken into account$^{25}$,the formulae for
neutrino masses are:(assuming generation mixings to be small)
$$m_{\nu_e}=(.05){m_u^2\over{M_N}} \;\eqno(23a)$$

$$m_{\nu_{\mu}}=(.07){m_c^2\over{M_N}} \;\eqno(23b)$$

$$m_{\nu_{\tau}}=(.18){m_t^2\over{M_N}} \;\eqno(23c)$$

In order to find the neutrino masses,we need to know $M_N$
which is given by $M_N=(f/g)M_R$.It can be argued on the
basis of vacuum stability $^{26}$ that $f\leq{g}$.The mean
value of $M_R$ has been obtained from two loop analysis of
the two SO(10) models in ref.10 and 27. For case (A) we have
$M_C^0\simeq{10^{11.5}}$GeV,whereas for model(B),we have
$M_R^0\simeq{10^{9}}$GeV.The uncertainty in the exponent
due to the error in $\alpha_{s}$  and $sin^2\theta_W$
is about $\pm.025$ in model(A) and $\pm.18$ in model(B).
Including this and the threshold uncertainties,
we find the minimum value of neutrino masses (corresponding
to f=g ) to range between the following values:

Model(A):$m_{\nu_e}=6\times{10^{-12}}eV-2\times{10^{-7}}eV$
;$m_{\nu_{\mu}}=4\times{10^{-7}}eV-10^{-2}eV$;
$m_{\nu_{\tau}}=.006eV-180eV$.

Model(B):$m_{\nu_e}=10^{-7}eV-10^{-6}eV$;$m_{\nu_{\mu}}=
10^{-2}eV-10^{-1}eV$ and $m_{\nu_{\tau}}=100eV-1 keV$.

We see that if the adiabatic solution is ruled out as is
currently believed,then model(B) will be ruled out by the solar
neutrino experiments and only model(A) will be acceptable.
This is an important result in our opinion since there
are no other uncertainties one can hide behind to save this
model.

{\bf VI.Higgs Boson Related Uncertainty in Proton Decay and
Stability of the Threshold calculations:}

In this section ,we discuss two questions:i)the uncertainty
in the predictions for proton decay and ii) the effect of
adding extra Higgs bosons to a GUT theory on the above
calculations.First,
we discuss the predictions for proton in the two SO(10)
models under discussion.Again using the results of ref.10 and 27
we find that for $\alpha_{s}=.11$,the value of $M_U=10^{15.8}$ GeV.
The uncertainty in $\alpha_{s}$,leads to an uncertainty of order
$10^{\pm.22}$ (see eq(8) and (22))
 multiplying the above value.We predict the proton
life-time for the model(A) to be $\tau_p=1.6\times10^{35\pm.7
\pm.9^{+3.2}_{-6.8}}$ years.For the model(B),we find,
$\tau_p=1.6\times10^{35\pm.7\pm1.\pm.8}$ years.

Let us now turn to the question of the stability of our results.
It is sometimes stated that any additional Higgs multiplet added
to a GUT model will add to the already existing uncertainty.However,
in a recent paper$^{28}$,it has been shown by one of the authors
that if the additional Higgs multiplet does have a vev or has vev
in a gauge direction which has been broken by a Higgs field with
the same representation content,then threshold effects from such
multiplets always cancel in $sin^2\theta_W$ or the intermediate
scales.This lends a degree of stability to the above calculations.

{\bf VII. Conclusions:}

In conclusion,we have presented a detailed analysis of the threshold
effects due to the unknown masses of the Higgs bosons and shown their
effect on the numerical predictions for the values of the unification
and the intermediate scales in two SO(10) models with a two step breaking.
To the best of our knowledge,this is the first time such an analysis
has been carried out for the SO(10) models.An interesting outcome
of this analysis is that the nonadiabatic MSW solution to the solar
neutrino puzzle is inconsistent with the model(B) which has an
intermediate symmetry $SU(2)_LXSU(2)_RXU(1)_{B-L}XSU(3)_c$.

\begin{center}
{\bf Acknowledgement}\\
\end{center}

This work is supported by a grant from the national Science Foundation.
One of the authors (M.K.P.) would like to thank the Council of
International Exchange of Scholars,Washington D.C.,and the University
Grants Commission,New Delhi for financial support.

\newpage

\newcounter{000}
\centerline{\bf References}
\begin{list}{[~\arabic{000}~]}
{\usecounter{000}\labelwidth=1cm\labelsep=.5cm}
\item J.C.Pati and A.Salam,{\it Phys.Rev.D$\,$}{\bf 10},275 (1974);\newline
H.Georgi and S.L.Glashow,{\it Phys.Rev.Lett.}{\bf 32},438 (1974)
\item H.Georgi,{\it Particles and Fields} ed.C.E.Carlson,A.I.P.,(1975)\newline
H.Fritzsch and P.Minkowski,{\it Ann.Phy.}{\bf 93},193 (1975).
\item S.Dimopoulos and H.Georgi,{\it Nucl.Phys.}{\bf B193},150 (1981).
\item For a recent summary of LEP data,see T.Hebbekar,Review talk
 at the LEP-HEP conference,Aachen preprint PITHA 91/17.\newline
\item P.Langacker and M.Luo,{\it Phys.Rev.D$\,$}{\bf 44},817 (1991)\newline
 U.Amaldi,W.de Boer and H.Furstenau,{\it Phys.Lett.B$\,$}{\bf 260},447(1991)
\newline
J.Ellis,S.Kelly and D.V.Nanopoulos,{\it Phys.Lett.B$\,$}{\bf 260},131 (1991).
\item J.C.Pati and A.Salam,{\it Phys.Rev.D$\,$}{\bf 10},275 (1974)\newline
R.N.Mohapatra and J.C.Pati,{\it Phys.Rev.D$\,$}{\bf 11},566,2558 (1975)\newline
G.Senjanovic andR.N.Mohapatra,{\it Phys.Rev.D$\,$}{\bf 12},1502 (1975)\newline
\item D.Chang,R.N.Mohapatra and M.K.Parida, {\it Phys. Rev. Lett.}
{\bf 52}, 1072 (1984);\newline
{\it Phys.Rev.D$\,$}{\bf 30},1052 (1984)
\item M.Gell-Mann,P.Ramond and R.Slansky,in {\it Supergravity}
(edited by D.Freedman {\it et.al}),North Holland,(1980);T.Yanagida,
{\it Proceedings of the Workshop on the Baryon Number of the Universe}
(edited by O.Sawada {\it et.al} (KEK,1979);\newline
R.N.Mohapatra and G.Senjanovic,
{\it Phys.Rev.Lett.}{\bf 44},912 (1980).
\item D.Chang and R.N.Mohapatra,{\it Phys.Rev.D$\,$}{\bf 32},1248 (1985).
\item D.Chang,R.N.Mohapatra,J.Gipson,R.E.Marshak and M.K.Parida,
 {\it Phys. Rev. D$\,$}{\bf 31}, 1718 (1985).
\item G.Cook,K.T.Mahanthappa and M.Sher,{\it Phys.Lett.B$\,$}{\bf 90},398
(1980)\newline
L.Hall,{\it Nucl.Phys.B$\;$}{\bf 178},75 (1981)
\item V.Dixit and M.Sher,{\it Phys.Rev.D$\,$}{\bf 40},3765 (1989).\newline
M.K.Parida and C.C.Hazra,{\it Phys.Rev.D$\;$}{\bf 40},3074 (1989).
\item M.K.Parida and P.K.Patra,{\it Phys.Rev.Lett.} {\bf 66},858 (1991).
\item S.Weinberg,{\it Phys.Lett.}{\bf B91},51 (1980)\newline
L.Hall,{\it Nucl.Phys.B$\;$}{\bf 178},75 (1981).
\item R.N.Mohapatra and M.K.Parida,University of Maryland Preprint\newline
UMD-PP-92-170 (1992).
\item P.Langacker,University of Pennsylvania Preprint UPR-0492-T (1992).
\item R.N.Mohapatra and G.Senjanovic,{\it Phys.Rev.D$\,$}{\bf 27} 1601
(1983)\newline
F.Del Aguila and L.Ibanez ,{\it Nucl.Phys.B$\,$}{\bf 177} 60 (1981).
\item See R.Barbieri and L.Hall,{\it Phys.Rev Lett.}{\bf 68},752 (1992)\newline
for a similar assumption.
\item L.F.Li,{\it Phys.Rev.D$,$}{\it 9}, 1723 (1974)\newline
 D.Chang and A.Kumar,{\it Phys.Rev.D$\,$}{\bf 33},2695 (1986)\newline
S.Meljanac and D.Pottinger,{\it Phys.Rev.D$\,$}{\bf 34},1654 (1986)\newline
X.-G.He and S.Meljanac,{\it Phys.Rev.D$\,$}{\bf 40} 2098 (1989).\newline
J.Baseq,S.Meljanac and L.O'Raifeartaigh,{\it Phys.Rev.D$\;$}{\bf 39},
3110 (1989)\newline
\item H.A.Bethe and J.Bahcall,{\it Phys.Rev.Letters}{\bf 65},2233
(1990)\newline
V.Barger,R.J.Phillips and K.Whisnant,{\it Phys.Rev.D$\;$}{\bf 43},
431 (1991);\newline
T.K.Kuo and J.Pantaleone,{\it Phys.Rev.D$\;$}{\bf 41},297 (1990);\newline
P.I.Krastev,S.P.Mikheyev and A.Yu Smirnov,{\it INR-preprint}{\bf 0695};\newline
For a recent review,see A.Yu Smirnov,Invited Talk at LEP-HEP\newline
meeting,Valencia Preprint (1992).
\item R.Davis Jr.,Proceedings of 'NEUTRINO'88" ed.J.Schnepps et al\newline
p.158,World Scientific(1989).
\item K.S.Hirata et.al.{\it Phys.Rev.Letters} {\bf 65},1297 (1990)\newline
P.Casper et.al, {\it Phys.Rev.Letters} {\bf 66},2561 (1991)\newline
\item A.Abazov et.al. {\it Phys.Rev.Letters} {\bf 67},3332 (1991).
\item For a recent overview,see S.F.Tuan {\it Int.Jour.Mod.Phys.A} (to appear)
\newline
\item S.Bludman,D.Kennedy and P.Langacker,Pennsylvania Preprint UPR-
0443T (1991).
\item R.N.Mohapatra,{\it Phys.Rev.D$\,$}{\bf 34},909 (1986).
\item F.Buccella,G.Miele,L.Rosa,P.Santorelli and T.Tuzi,{\it Phys.Lett.}
{\bf 233B},178 (1989);
M.K.Parida,P.K.Patra and C.C.Hazra,{\it Phys.Rev.D$\,$}{\bf 43},2351
(1991).\newline
N.Deshpande,E.Keith and P.B.Pal,Oregon Preprint OITS-484 (1992)
\item R.N.Mohapatra,Unversity of Maryland Preprint UMD-PP-92-170.

\end{list}
\newpage
\begin{center}
{\bf Table Caption}
\end{center}

Table I.The Higgs Bosons at Mass scale $M_U$.

Table II.The Higgs bosons with masses at $M_C$;
     The numbers within the parenthesis refer to the
     representation content under $SU(2)_LXU(1)_YXSU(3)_C$
  The multiplet $\phi$arises from the $\phi(2,2,0)$ and
  the R-multiplets arise from the multiplet $\Delta_R(1,3,\bar{10})$.

Table III.In this table, we present our results for the threshold
uncertainties in the intermediate scale and the unification scale
for different values of $e^\eta$.The first four lines correspond to
the case where the uncertainty in $M_I$ is maximized whereas the
last four lines correspond to the case where the uncertainty in

Table IV.Higgs bosons with masses of order $M_U$ .

\begin{center}
{\bf Table I}
\end{center}
\begin{tabular}{|c||c|}\hline
   SO(10) Representation  &  $G_{224}$ submultiplet \\  \hline
        {10}              & H(1,1,6)          \\    \hline
        {126}      & $\zeta_0$(2,2,15),S(1,1,6),$\Delta_L$(3,1,10)\\ \hline
        {210}           & $\Sigma_L$(3,1,15),$\Sigma_R$(1,3,15),\\
                        &$\zeta_1$(2,2,10),$\zeta_2(2,2,\bar{10})$ \\
                        &$\zeta_3$(1,1,15),S'(1,1,1,),U(2,2,6) \\  \hline
\end{tabular}
\newpage
\begin{center}
{\bf Table II}
\end{center}
\begin{tabular}{|c||c|}\hline
  SO(10) representation &  $G_{213}$ submultiplet at $M_C$ \\  \hline
          {10}          &  $\phi(2,-\sqrt{3/5}{1\over2},1)$ \\  \hline
          {126}   & $R_1(1,\sqrt{3/5}{1\over3},\bar{3})$,  \\
                  & $R_2(1,\sqrt{3/5}{1\over3},\bar{6}$  \\
                  & $R_3(1,-\sqrt{3/5}{2\over3},\bar{6})$ \\
                  & $R_4(1,-\sqrt{3/5}{4\over3},\bar{3})$  \\
                  & $R_5(1,\sqrt{3/5}{4\over3},\bar{6})$  \\
                  & $R_6(1,2\sqrt{3/5},1)$    \\    \hline
\end{tabular}

\begin{center}
{\bf Table III}\\
\end{center}
\begin{tabular}{|c||c||c||c|}\hline
Symmetry Breaking chain& $M_H/M_U$ or$M_H/M_I$& $M_I/M^{0}_I$ & $M_U/M^{0}_U$\\
   \hline

$G_{224}$  &  1/30 to 30   &  $10^{ ^{+4}_{-2.1}}$ & $10^{^{+1.2}_{-2.5}}$ \\
$G_{2213}$ &           &  $10^{^{+.9}_{-.4}}$ & $10^{^{+.1}_{-.2}}$\\ \hline
$G_{224}$  &  1/10 to 10   &  $10^{^{+2.7}_{-1.4}}$  & $10^{^{+.8}_{-1.7}}$ \\
$G_{2213}$ &            &  $10^{^{+.6}_{-.3}}$  & $10^{\pm.1}$  \\  \hline
$G_{224}$  & 1/30 to30     & $10^{^{+4.2}_{-2.2}}$ & $10^{^{+1.2}_{-2.5}}$ \\
$G_{2213}$ &               & $10^{^{+.5}_{-.2}}$ & $10^{\pm.2} $ \\  \hline
$G_{224}$  & 1/10 to 10    & $10^{^{+2.8}_{-1.5}}$ & $10^{^{+.8}_{-1.7}}$ \\
$G_{2213}$ &               & $10^{^{+.3}_{-0}}$   & $10^{\pm.2}  $  \\  \hline

\end{tabular}

\newpage
\begin{center}
{\bf Table IV}
\end{center}
\begin{tabular}{|c||c|}\hline
  SO(10) representation & $G_{2213}$ content of the heavy boson\\  \hline
        {10}    &  $T_1(1,1,{1\over3}\sqrt{3/2},3)$  \\
                &  $T_2(1,1,-{1\over3}\sqrt{3/2},\bar{3})$\\  \hline
        {126}   & $H_{1L}(3,1,-{1\over3}\sqrt{3/2},6)$ \\
                & $H_{1R}(1,3,+{1\over3}\sqrt{3/2},\bar{6})$\\
                & $H_{2L}(3,1,-{1\over3}\sqrt{3/2},3)$\\
                & $H_{2R}(1,3,+{1\over3}\sqrt{3/2},\bar{3})$\\
                & $H_{3}(2,2,-{2\over3}\sqrt{3/2},3)$ \\
                & $H_4(2,2,+{2\over3}\sqrt{3/2},\bar{3})$ \\
                & $H_5(2,2,0,8)$,$H_6(2,2,0,1)$ \\
                & $H_7(1,1,{1\over3}\sqrt{3/2},3)$ \\
                & $H_8(1,1,-{1\over3}\sqrt{3/2},\bar{3})$ \\ \hline
        {45}    & $S_1(1,1,0,8),S_2(3,1,0,1),S_3(1,3,0,1)$ \\ \hline
\end{tabular}
\end{document}